# Diffusion posterior sampling enables zero shot kilometre scale wind forecasting over complex terrain

Yujiang Cai[1,2], Ya Wang[1,2], Shuanglei Feng[3], Bo Wang[3], Yiming Liu[1,2], Hui Yuan[4], Xinyi Sun[1,2], Haijie Li[1,2] and Shenming Fu[1,2]

[1] *China Key Laboratory of Earth System Numerical Modeling and Application, Institute of Atmospheric Physics, Chinese Academy of Sciences, Beijing 100029, China.*

[2] *University of Chinese Academy of Sciences, Beijing 100049, China.*

[3] *National Key Laboratory of Renewable Energy Grid-Integration, China Electric Power Research Institute, Beijing 100192, China.*

[4] *State Grid Shanxi Electric Power Research Institute, Taiyuan 030001, China.*

Corresponding authors: Dr. Ya Wang (wangya@mail.iap.ac.cn)

Dr. Shenming Fu (fusm@mail.iap.ac.cn)

## Abstract

Reliable prediction of near surface wind over complex terrain is limited by the mismatch between the spatial resolution of operational forecasts and terrain controlled local wind variability. Here, we develop KiloGen, a diffusion posterior sampling framework for kilometre scale wind forecast enhancement. KiloGen learns a high resolution vector wind prior from Weather Research and Forecasting (WRF) model simulations and constrains posterior sampling with 25 km forecasts from the European Centre for Medium Range Weather Forecasts (ECMWF) at inference time. This formulation avoids paired ECMWF and WRF training samples and an explicitly learned mapping from coarse to fine resolution. Applied over Shanxi, China, a region with complex mountainous terrain, KiloGen reconstructs terrain organized wind structures and restores high wavenumber variability while retaining the large scale evolution of the operational forecast. Station verification shows that KiloGen achieves the lowest overall wind speed root mean square error (RMSE) among the evaluated products, with larger benefits at elevated and topographically complex sites. The improvement is strongest under strong wind conditions, reducing RMSE by approximately 10% for observed winds above 20 m $s^{-1}$. Across 13 distinct strong wind events, KiloGen improves upon the 0.25° ECMWF forecast in all cases and outperforms the 0.1° ECMWF forecast in most cases. These results show that diffusion posterior sampling provides an effective approach for terrain aware kilometre scale wind forecast enhancement.

## Introduction

Accurate near surface wind forecasts are vital for renewable energy operations, aviation safety, and disaster preparedness, yet they are particularly hard to obtain in mountainous regions. Near surface wind forecasts over complex terrain are limited by a persistent scale mismatch. Operational numerical weather prediction can provide skilful large scale circulation and temporal evolution, whereas local wind speed and direction are strongly modulated by ridges, valleys, slope breaks and basin edges at kilometre and subkilometre scales.[1-7] These terrain controlled processes create sharp spatial contrasts in wind speed and direction that are important for strong wind warnings, local risk assessment and wind energy operations, but are inevitably smoothed or displaced in coarse forecast fields.[8-10] This limitation is particularly consequential in regions where large operational forecast errors can overlap with dense wind-turbine deployment, because unresolved terrain scale wind variability can directly affect wind energy assessment, power production and operational risk management.[2,3]

Several approaches have been developed to bridge this scale gap. Interpolation and remapping increase grid density but cannot recover spatial variance absent from the original forecast.[11,12] Statistical correction can improve marginal wind speed distributions but does not explicitly reconstruct flow dependent vector wind structures.[12,13] Regional dynamical downscaling with the Weather Research and Forecasting (WRF) model resolves terrain effects more directly, although repeated kilometre scale integrations are computationally demanding.[9,14,15] Supervised deep learning downscaling can learn coarse to fine mappings, but usually requires paired training samples and may remain tied to a particular input product, resolution or training configuration.[16-21]

Recent advances in diffusion based inverse modelling provide an alternative to directly learning a fixed coarse to fine mapping.[22-24] Denoising diffusion restoration and diffusion posterior sampling combine a pretrained generative prior with measurement constraints during reverse diffusion, enabling reconstruction under different observation operators without retraining the prior model.[25,26] In the atmospheric domain, this paradigm has been pioneered by frameworks like SAGE alongside related formulations for atmospheric reconstruction, climate downscaling, and partial observation state inference.[27-31] However, existing studies have primarily focused on general image restoration, climate downscaling, sparse observation reconstruction and partial weather state inference. How a regional terrain resolving wind prior can be integrated with forecast time operational guidance to improve near surface vector winds over complex terrain remains insufficiently explored.

Here we develop KiloGen, a diffusion posterior sampling method for kilometre scale 10 m vector wind reconstruction.[22-31] The method separates two sources of information that are commonly entangled in deterministic downscaling. A diffusion model learns a reusable high resolution prior from WRF simulated 10 m zonal wind (U10) and 10 m meridional wind (V10) fields, where terrain organized wind structures are explicitly represented. During inference, 0.25° forecasts from the European Centre for Medium Range Weather Forecasts (ECMWF), hereafter referred to as EC 0.25°, constrain the posterior samples at the spatial scales resolved by the operational model. KiloGen therefore does not learn a direct EC to WRF mapping. In this study, the term zero shot refers specifically to the absence of paired ECMWF and WRF training samples and ECMWF specific coarse to fine retraining within the study domain, rather than to cross region or cross season transfer.

# Results

## Operational wind forecast errors increase with terrain complexity

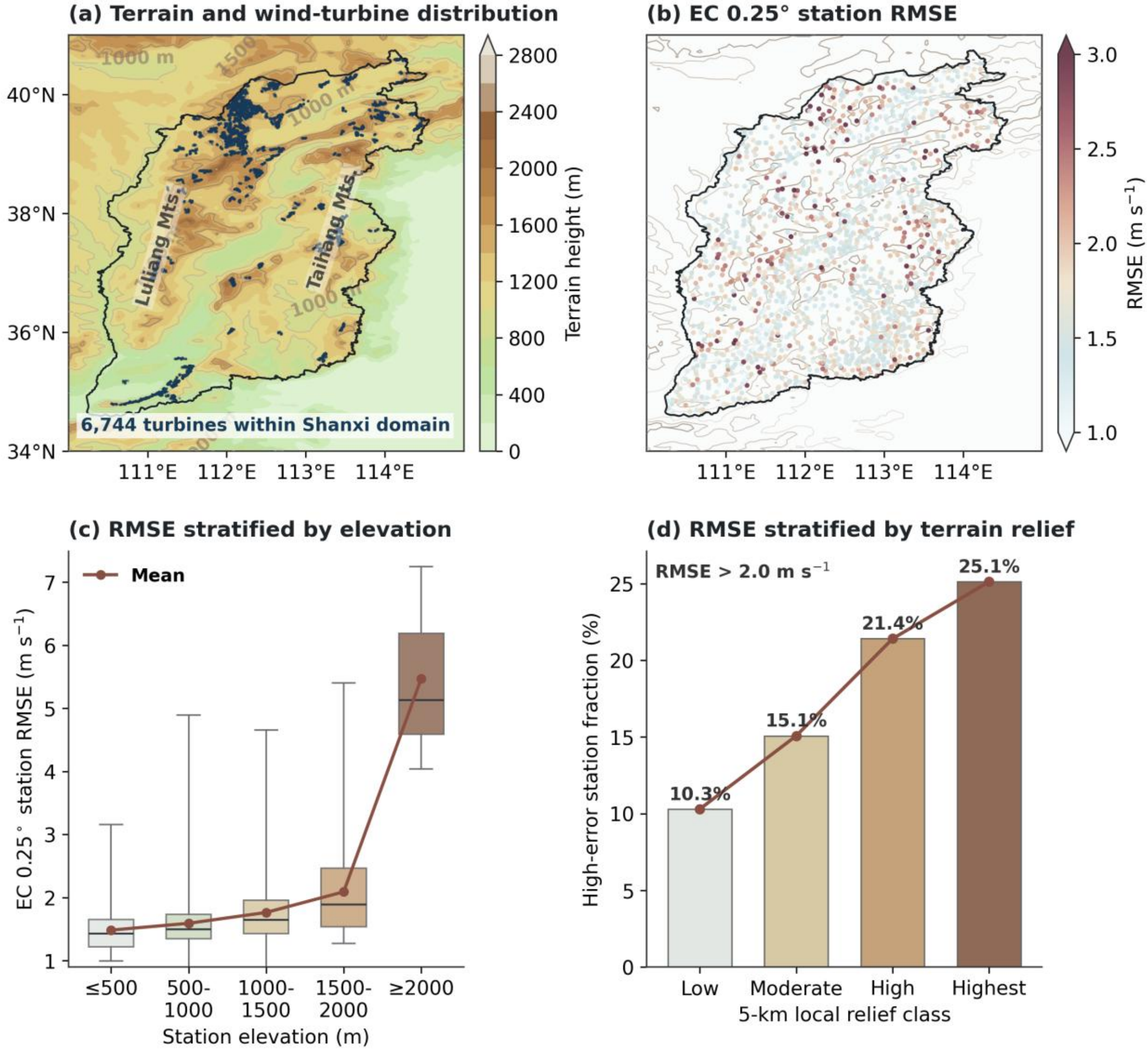

**Fig.** 1 | **Terrain dependence of operational wind forecast errors**. a, Terrain height and wind-turbine distribution over Shanxi and the surrounding region. b, Station root mean square error (RMSE) of the EC 0.25° forecast for 10 m wind speed. c, EC 0.25° station RMSE stratified by station elevation. d, Fraction of high error stations, defined as RMSE > 2.0 m s$^{-1}$, grouped by 5 km local terrain relief class.

The Shanxi region provides a suitable testbed for kilometre scale wind reconstruction because terrain height and exposure vary sharply over short distances (Fig. 1a). Within the study domain, 6,744 wind-turbines are distributed across plateau margins, mountain ridges, basins and valley corridors, where local acceleration, sheltering and channeling can strongly influence near surface winds. These structures occur at scales that are only weakly represented on a 0.25° forecast grid.

This scale mismatch is reflected in the EC 0.25° station errors (Fig. 1b-d). The domain mean RMSE is 1.75 m $s^{-1}$, with a 90th percentile value of 2.28 m $s^{-1}$. Errors increase with station elevation and become more frequent as 5 km local terrain relief increases. The high error fraction rises from 10.3% in the lowest relief class to 25.1% in the highest relief class. These results show that the limitations of the coarse forecast are systematically concentrated in terrain sensitive locations, motivating a reconstruction strategy that retains the large scale forecast while restoring unresolved local wind variability.

## KiloGen combines a high resolution wind prior with an operational forecast constraint

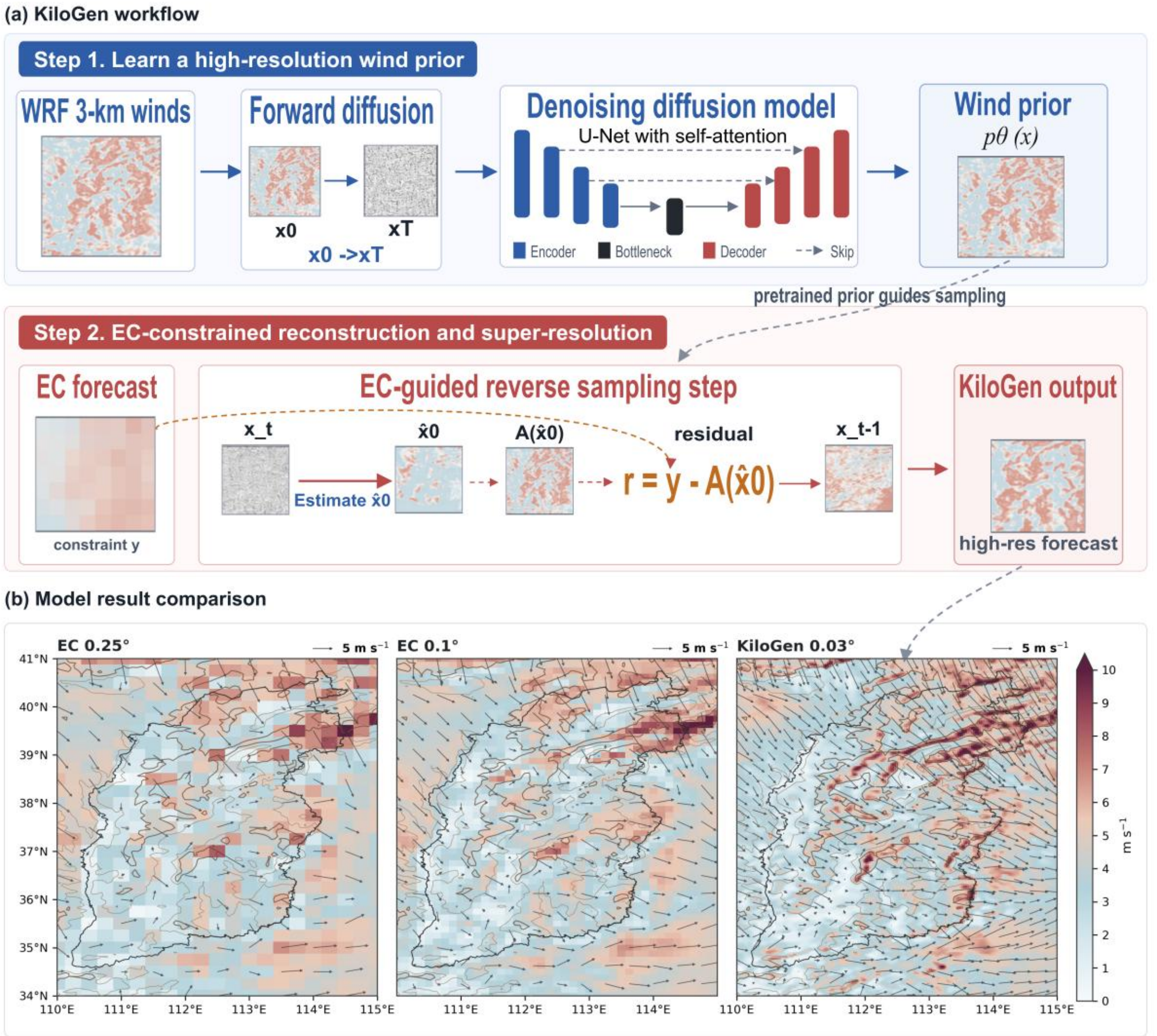

**Fig.** 2 | **KiloGen workflow and prior constrained wind reconstruction**. a, Schematic of the KiloGen workflow. A high resolution vector wind prior is learned from WRF simulations and combined with an EC 0.25° forecast during posterior sampling. b, Example wind field comparison among EC 0.25°, 0.1° ECMWF forecast (EC 0.1°) and KiloGen 0.03° over the same domain. Shading denotes 10 m wind speed, and arrows denote horizontal 10 m wind vectors. The colour scale is in m $s^{-1}$, and the reference vector indicates 5 m $s^{-1}$.

Figure 2 summarizes the prior constraint formulation. The WRF trained diffusion model represents plausible kilometre scale U10 and V10 structures, whereas the EC 0.25° forecast provides the forecast time large scale constraint. During reverse diffusion, the generated field is repeatedly mapped to the EC grid and adjusted toward the operational forecast, allowing unresolved local variability to be reconstructed without discarding the regional flow background. In the example shown, EC 0.25° captures the broad wind pattern but smooths local gradients, while EC 0.1° adds limited spatial detail. KiloGen retains the regional circulation while producing sharper and more continuous terrain following wind structures. The reconstructed field is best interpreted as a constrained enhancement of the operational forecast rather than as a free running high resolution prediction.

## The learned prior represents kilometre scale wind variability

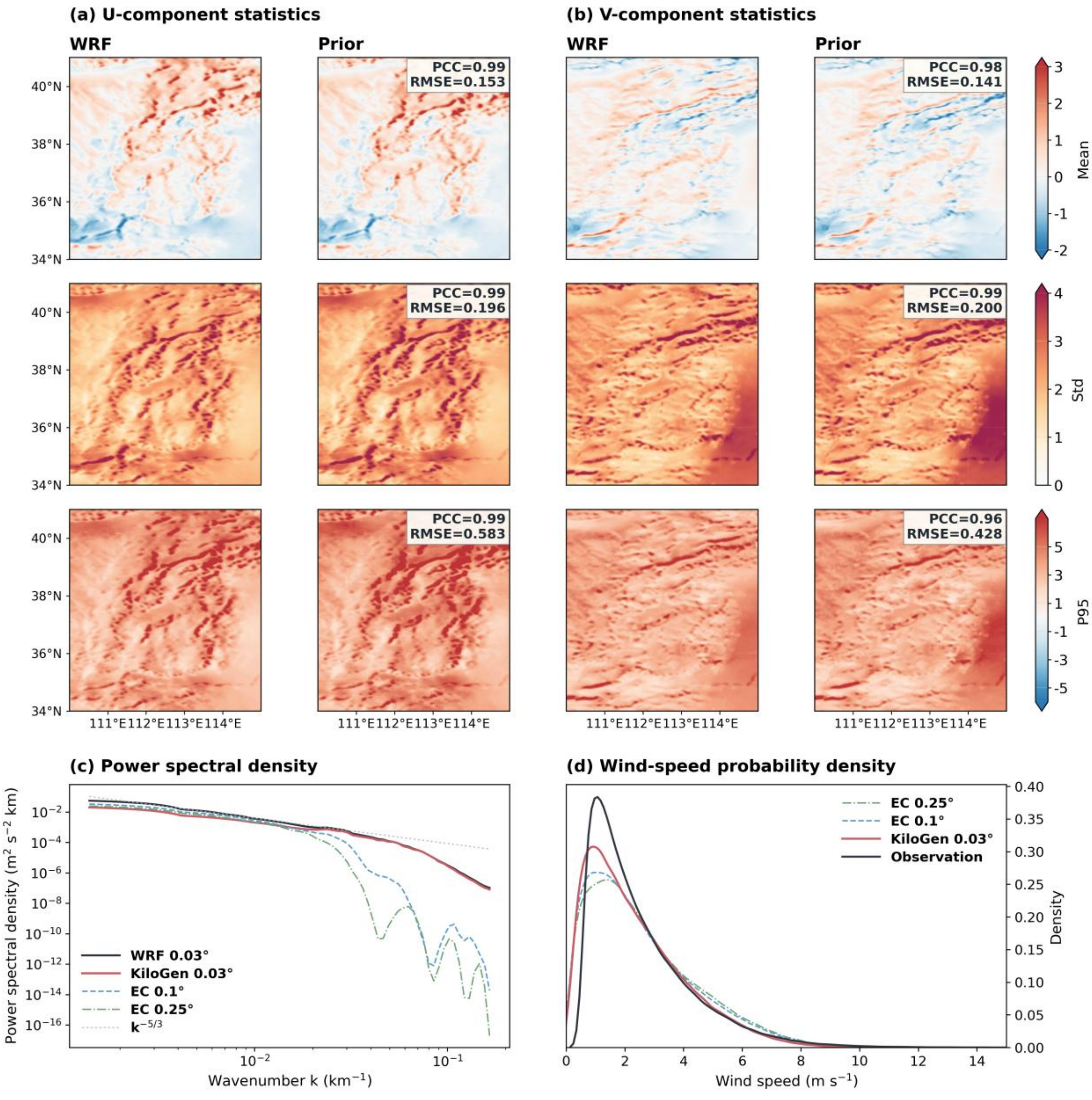

**Fig.** 3 | **Validation of the WRF based wind prior and reconstructed variability**. a,b, Spatial maps of the mean, standard deviation and 95th percentile of 10 m zonal wind (U10; a) and 10 m meridional wind (V10; b) from WRF and generated prior samples. Pattern correlation coefficient (PCC) and RMSE between the WRF statistics and generated prior statistics are shown in each prior panel. c, Power spectral density of 10 m wind speed fields from WRF 0.03°, KiloGen 0.03°, EC 0.1° and EC 0.25°. d, Station matched probability density distributions of 10 m wind speed from observations, EC 0.25°, EC 0.1° and KiloGen 0.03°. Colour bars in a and b denote the corresponding wind component statistics.

Figure 3 tests whether the trained diffusion model reproduces the statistical structure of the WRF based high resolution prior. We generate 1,000 samples from the trained model and compare their statistics with those of the WRF training data. The generated samples closely match the WRF mean, standard deviation and 95th percentile fields for both U10 and V10 (Fig. 3a,b), with pattern correlations of 0.96 to 0.99. They also retain terrain aligned bands and localized wind maxima over mountain ranges, plateau margins and transition zones. These diagnostics show that the diffusion model captures organized spatial variability in the WRF simulations, not just domain scale gradients.

The spectral and station distribution diagnostics show how this learned variability affects the posterior reconstructions (Fig. 3c,d). EC 0.25° and EC 0.1° lose high wavenumber power relative to WRF 0.03°, consistent with coarse grid smoothing. KiloGen retains substantially more high wavenumber variance while remaining constrained by the ECMWF forecast at resolved scales. At station locations, its wind speed distribution is also closer to the observed distribution over the main wind speed range. Together, these results indicate that the posterior reconstruction adds WRF consistent fine scale variability that is relevant to observed station winds.

## Prior constrained reconstruction improves station wind forecasts

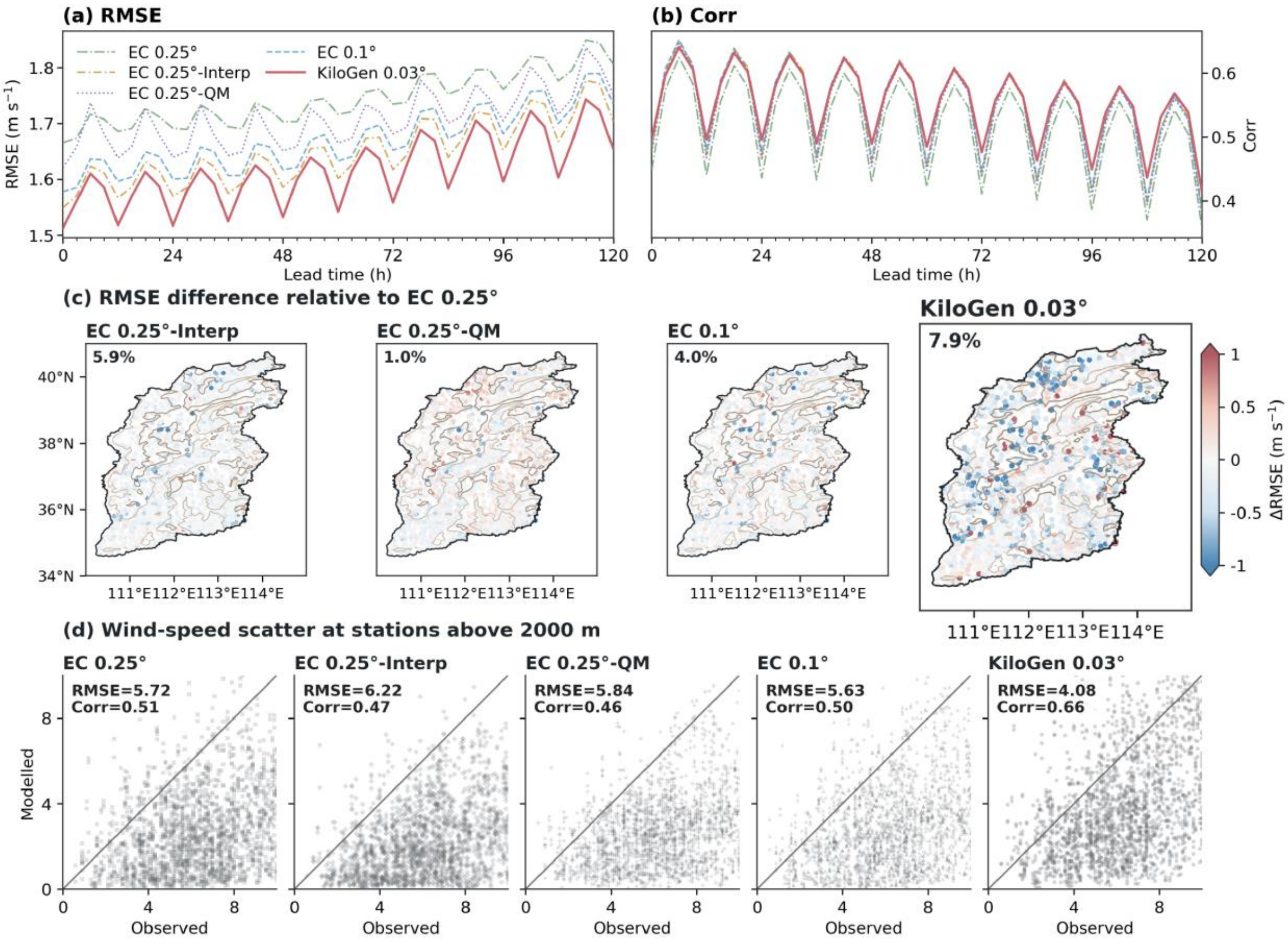

Fig. 4 | Station scale forecast skill of KiloGen. a, Wind speed RMSE as a function of forecast lead time from 0 to 120 h for EC 0.25°, EC 0.25°-Interp, EC 0.25°-QM, EC 0.1° and KiloGen 0.03°. b, Pearson correlation coefficient as a function of forecast lead time for the same products. c, Station RMSE differences relative to EC 0.25°. Blue indicates RMSE reduction, and red indicates RMSE increase. The percentage in each panel denotes the domain mean RMSE reduction relative to EC 0.25°. d, Observed versus predicted wind speed at stations above 2000 m elevation. The grey diagonal line denotes the 1:1 line.

Beyond reconstructing more detailed and physically plausible kilometre scale wind fields, the key question is whether the added spatial variability translates into improved station scale forecast skill. We evaluate KiloGen using approximately 1,000 quality controlled meteorological stations available at each valid time across Shanxi. The comparison includes the EC 0.25° forecast used as the large scale constraint, the higher resolution EC 0.1° forecast as a stronger operational baseline, and two postprocessing baselines derived from EC 0.25°. EC 0.25°-Interp bilinearly interpolates the 0.25° forecast to the 0.03° target grid, whereas EC 0.25°-QM further applies quantile mapping to align its wind speed distribution with that of the kilometre scale WRF training data.

The lead-time verification shows that KiloGen provides the most consistent improvement in station scale wind speed forecasts over the 0 to 120 h forecast window (Fig. 4a and b). Relative to the constraining EC 0.25° forecast, KiloGen achieves the largest domain mean RMSE reduction among all evaluated products, with a 7.9% decrease, compared with 5.9% for EC 0.25°-Interp, 1.0% for EC 0.25°-QM and 4.0% for EC 0.1°. The RMSE advantage persists across most lead times, including both the low error phases and the daily error peaks, indicating that the reconstruction does not simply smooth the forecast but improves the time dependent station match. KiloGen also maintains slightly higher correlation than EC 0.25° and is comparable to or higher than EC 0.1° over most lead times, suggesting that the RMSE reduction is achieved while largely preserving the temporal evolution inherited from ECMWF.

The spatial diagnostics show that the improvement is not spatially uniform, but is concentrated in the locations where coarse forecasts are least representative (Fig. 4c and d). The largest RMSE reductions occur near mountain ranges, terrain transition zones and other areas of pronounced local complexity. At stations above 2,000 m, KiloGen reduces RMSE from 5.72 to 4.08 m $s^{-1}$ and increases the correlation coefficient from 0.51 to 0.66 (Fig. 4d). The principal value of the reconstruction therefore lies not only in its domain mean improvement, but also in its substantially larger benefit in terrain sensitive environments where unresolved local wind variability is most consequential.

Additional wind direction verification and mean absolute error (MAE) diagnostics provide complementary evidence that the improvements are not limited to wind speed RMSE alone (Supplementary Figs. 1 and 2).

## KiloGen reduces upper tail wind speed errors

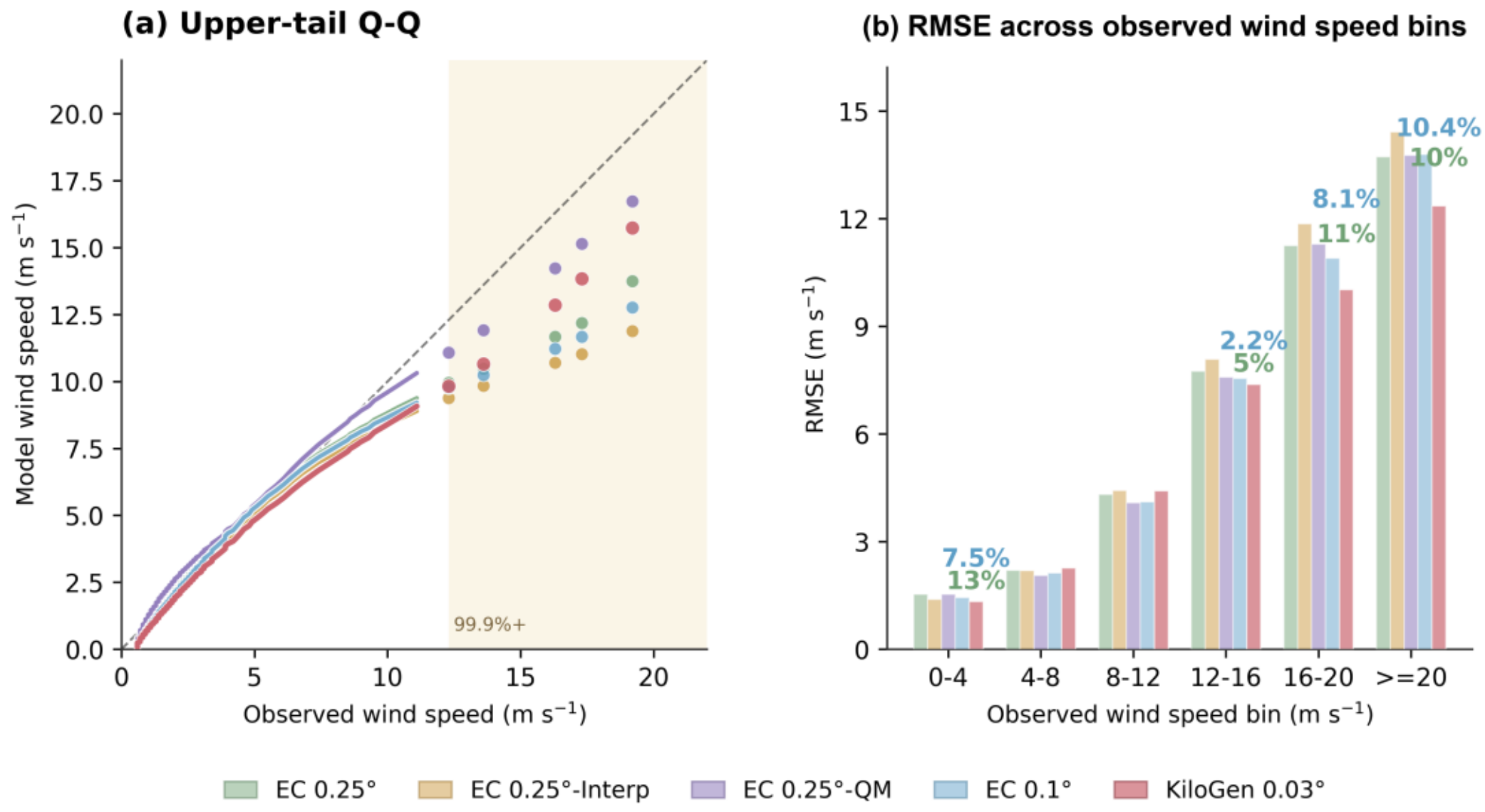

Fig. 5 | Forecast performance across different wind speed regimes. a, Upper tail quantile-quantile (Q-Q) plot of station matched 10 m wind speeds for EC 0.25°, EC 0.25°-Interp, EC 0.25°-QM, EC 0.1° and KiloGen 0.03° against observations. The shaded region marks the upper tail range above the 99.9th percentile of observed wind speed. b, RMSE across observed wind speed bins. Percentages above selected bars indicate KiloGen's relative RMSE reduction against EC 0.25° and EC 0.1°; green labels are relative to EC 0.25°, and blue labels are relative to EC 0.1°.

We next examine how forecast performance varies across the observed wind speed distribution. The Q-Q analysis shows that the evaluated products increasingly underestimate wind speed toward the upper end of the distribution (Fig. 5a). Above the 99.9th percentile, KiloGen is closer to the observed quantiles than EC 0.25°, EC 0.1° and the interpolation baseline. When observed wind speeds approach 20 m s$^{-1}$, the EC forecasts typically remain near 12 to 14 m s$^{-1}$, whereas KiloGen reconstructs values of approximately 16 m s$^{-1}$. Although the highest observed winds remain underestimated, KiloGen substantially reduces the amplitude deficit. Quantile mapping more directly adjusts the marginal distribution, but does not produce comparable reductions in station RMSE (Fig. 4a), suggesting that distributional correction alone cannot fully explain the improvement.

The RMSE across observed wind speed bins provides a more detailed assessment over the full wind speed range (Fig. 5b). Differences among the five products are relatively small under low and moderate wind conditions, whereas the advantage of KiloGen becomes progressively clearer as observed wind speed increases. In the bins from 16 to 20 m $s^{-1}$ and above 20 m $s^{-1}$, KiloGen reduces RMSE by 8.1% to 10.4% relative to EC 0.1° and by approximately 10% to 11% relative to EC 0.25°. These results indicate that KiloGen improves forecast performance across the wind speed distribution, with the largest benefits occurring at higher wind speeds, where coarse resolution forecasts exhibit the strongest amplitude underestimation.

## Strong wind events show consistent event scale improvements

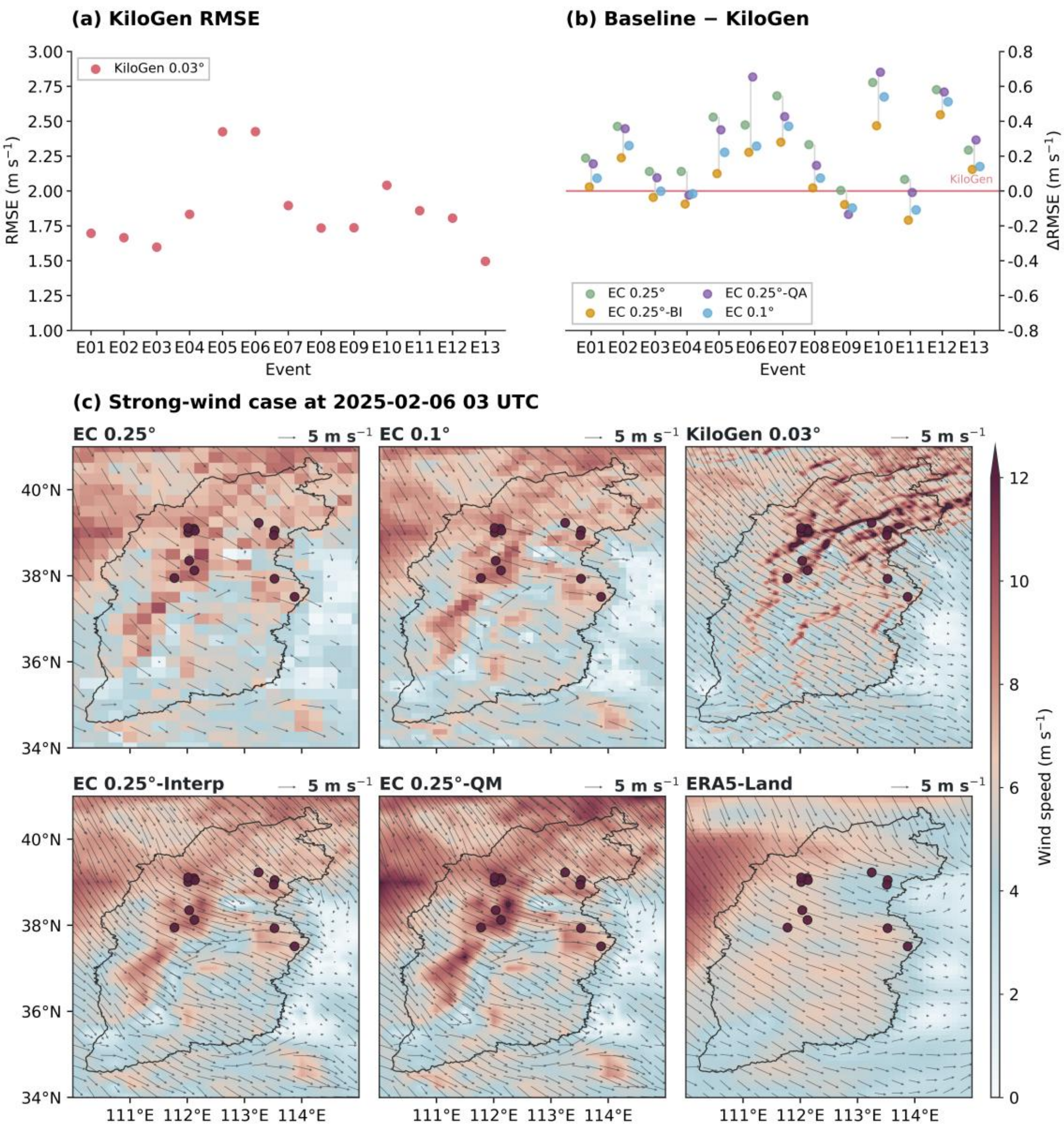

Fig. 6 | Event scale strong wind evaluation and the E06 case. a, Event wise RMSE of KiloGen 0.03° for 13 distinct strong wind events from January to March 2025. b, RMSE difference between each baseline and KiloGen for the 13 events. Positive values indicate lower RMSE for KiloGen. Baselines include EC 0.25°, EC 0.25°-Interp, EC 0.25°-QM, EC 0.1° and ERA5-Land. c, Wind speed and 10 m wind vectors for the E06 case at 03 UTC on 6 February 2025, initialized at 00 UTC on 5 February 2025. Station markers indicate observed strong wind sites.

We next focus on strong wind events. Hourly station observations from January to March 2025 were grouped into events separated by at least 24 h. An event was retained when at least three stations recorded wind speeds of 12 m s$^{-1}$ or greater, and the peak hour of each eligible event was used for evaluation. This procedure identifies 13 distinct strong wind events. Across these events, KiloGen achieves a mean RMSE of 1.86 m $s^{-1}$, compared with 2.16 m $s^{-1}$ for EC 0.25°, 1.97 m $s^{-1}$ for EC 0.25°-Interp, 2.13 m $s^{-1}$ for EC 0.25°-QM and 2.03 m $s^{-1}$ for EC 0.1° (Fig. 6a and b). KiloGen produces lower RMSE than EC 0.25° in all 13 events and lower RMSE than EC 0.1° in 9 of the 13 events. This consistent advantage across events indicates that the improvement is not dominated by one selected case.

The E06 event illustrates how this improvement appears in the spatial wind field (Fig. 6c). At 03 UTC on 6 February 2025, 12 stations recorded wind speeds above 12 m $s^{-1}$. EC 0.25° and EC 0.1° capture the broad windy conditions, but their high wind regions remain relatively smooth and weaker than the station observations. Interpolation and quantile mapping modify the coarse forecast field, but do not recover similarly localized wind structures associated with the complex terrain. By contrast, KiloGen preserves the regional flow pattern while reconstructing more localized wind speed maxima along terrain following corridors, including ridge and valley transition zones where local acceleration and channeling are expected. These reconstructed structures are closer in amplitude to the station observations and provide a more physically plausible representation of the terrain modulated strong wind field.

Together, the event statistics and the E06 case show that KiloGen improves strong wind guidance not only by reducing station error, but also by restoring localized terrain aware wind structures that are smoothed in the operational forecasts. The event catalogue and additional strong wind reconstructions show that the selected events are temporally separated and that similar terrain aware reconstruction behaviour appears in other strong-wind cases, not only in the representative E06 case (Supplementary Figs. 3 and 4).

## Discussion

This study formulates kilometre scale near surface wind enhancement as a prior constrained posterior reconstruction problem. The high resolution regional wind distribution is learned independently from WRF simulations, while large scale circulation information from ECMWF is introduced only during posterior sampling. This separation avoids learning a fixed mapping from a specific coarse forecast product to a high resolution target. This design allows the EC 0.25° forecast to act as an inference time constraint without paired ECMWF and WRF training. Here, zero shot therefore refers to the absence of paired ECMWF and WRF training samples and ECMWF specific retraining within the Shanxi experiments, rather than demonstrated transfer to other forecast products, regions or seasons.

KiloGen achieves the lowest overall RMSE among the evaluated products, with the clearest improvements occurring over elevated terrain, near strong terrain gradients and at higher wind speeds. The analysis across observed wind speed bins shows that the reduction in wind speed underestimation becomes more pronounced as observed wind speed increases. The evaluation of 13 strong wind events further shows that the improvement is not dominated by a single selected case, with KiloGen outperforming EC 0.25° in every event and EC 0.1° in most events. These results indicate that the principal value of KiloGen lies in enhancing forecasts where unresolved terrain effects exert the greatest influence, rather than uniformly replacing operational numerical weather prediction.

Several limitations define the scope of these conclusions. The learned prior may inherit systematic biases from WRF, and the current verification is limited to Shanxi during winter and early spring 2025. Although the framework can in principle accept coarse inputs from different sources and resolutions, the present experiments do not directly demonstrate transfer across forecast products, regions or terrain regimes. The most extreme station winds also remain underestimated, suggesting that local exposure, gustiness, boundary layer stability and surface heterogeneity cannot be fully recovered from coarse forecast information alone. Reconstruction quality may also depend on the observation operator and the strength of posterior conditioning. Posterior members provide a sampling based indication of reconstruction uncertainty under the same EC 0.25° constraint, but because probabilistic calibration is not evaluated here, the spread is interpreted as posterior sampling variability rather than as a fully calibrated forecast uncertainty estimate (Supplementary Fig. 5).

Within these limits, KiloGen provides a flexible framework for kilometre scale wind forecast enhancement over complex terrain. By combining a WRF based high resolution wind prior with forecast time ECMWF constraints, it preserves the large scale forecast evolution while improving local wind amplitudes and station accuracy, particularly in terrain sensitive locations and under stronger wind conditions.

# Methods

## Overview: prior-constraint posterior reconstruction

KiloGen is built on diffusion posterior sampling, which combines a pretrained generative prior with a measurement constraint during the reverse diffusion process.[22-31] In this study, the pretrained prior is a WRF trained diffusion model that represents plausible kilometre scale U10 and V10 wind structures. The measurement constraint is provided by the EC 0.25° forecast, which supplies the forecast time large scale circulation and temporal evolution.

With the prior and constraint separated, KiloGen reconstructs kilometre scale wind fields without learning a fixed EC to WRF mapping. During inference, the sampler searches for high resolution wind fields that are plausible under the WRF based prior and remain anchored to the EC 0.25° forecast at operationally resolved scales. The output is a constrained refinement of the operational forecast, not a free running high resolution prediction. Posterior sampling produces multiple plausible reconstructions under the same EC 0.25° constraint. In this study, the ensemble mean is used as the deterministic KiloGen 0.03° product for verification.

## Study region and data

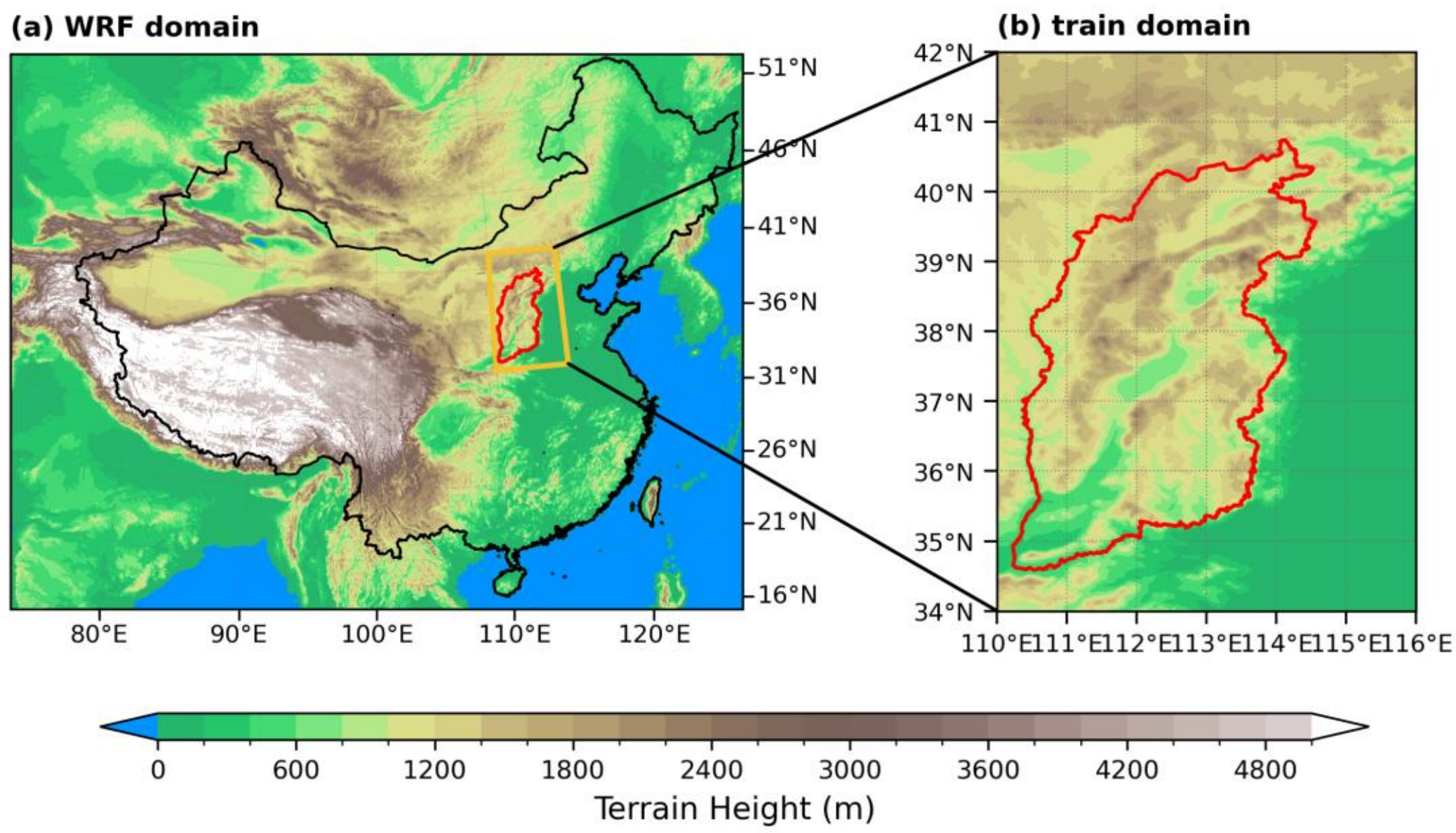

**Fig.** 7 | Study domain. a, Regional WRF simulation domain and terrain setting. b, Shanxi high resolution training and verification domain.

The study region is Shanxi, China, and the surrounding terrain that modulates near surface strong winds (Fig. 7b). The domain contains plateau margins, mountain ranges, basins and valley corridors, and includes a dense wind-turbine deployment for which kilometre scale wind guidance is practically relevant. The high resolution training and reconstruction domain spans 34.02-41.67°N and 110.01-115.74°E on a 256 × 192 grid with 0.03° horizontal spacing. The model variables are 10 m zonal wind (U10) and 10 m meridional wind (V10), and wind speed is calculated from the reconstructed vector wind as

$$\mathrm{WS} = \sqrt{(\mathrm{U10})^2 + (\mathrm{V10})^2} \quad (1)$$

Three data streams are used for distinct purposes in the experimental design. The high resolution wind prior is trained exclusively using 2,860 WRF U10 and V10 samples from January to December 2024, consisting of three hourly short lead outputs within the first 24 h of each available forecast cycle. EC 0.25° forecasts initialized at 00 and 12 UTC from January to March 2025 are used only as coarse constraints during posterior sampling, with forecast lead times from 0 to 120 h. EC 0.1° forecasts over the same period are used only as a higher resolution operational baseline. Hourly observations from approximately 1,000 quality controlled meteorological stations during January to March 2025 are used solely for independent verification. Although the evaluation period is limited to three months because dense station observations were available only for this period, the station network provides approximately 2.1 million station hour records before common sample filtering. The evaluation therefore provides extensive spatial and event sampling for winter and early spring conditions, while not establishing all season generality. Station observations are not used in prior training or posterior reconstruction, and no paired ECMWF and WRF samples are required during model training.

Terrain complexity in Fig. 1 is quantified using a station centred 5 km local relief metric derived from the WRF terrain field HGT_M. For each station, horizontal distance from the station to every terrain grid cell is approximated using 111 km per degree latitude and a cosine latitude correction for longitude. Terrain cells within a 5 km radius are selected, and local relief is defined as the 90th percentile minus the 10th percentile of terrain height within that neighbourhood. Stations are then grouped into quartiles of this 5 km local relief metric to define the Low, Moderate, High and Highest terrain complexity classes used for the high error risk analysis in Fig. 1d.

The prior training dataset contains 2,860 WRF samples from 1 January to 31 December 2024. The retained WRF samples are 3 hourly short lead outputs at 00, 03, 06, 09, 12, 15, 18 and 21 UTC for each available day, which reduces the influence of accumulated regional forecast errors. Each sample contains U10 and V10 fields cropped to the 256 by 192 target domain. The WRF production system uses ERA5 as the initial and lateral boundary condition and runs daily cycling forecasts with two active nested domains.[10] The outer and inner domains have horizontal grid spacings of 15 km and 3 km, respectively, with 37 vertical levels, 3 hourly lateral boundary updates and 91 h forecast integrations. The model is nonhydrostatic and uses adaptive time stepping.[9] The main physics configuration includes Thompson microphysics, the Rapid Radiative Transfer Model for General circulation models (RRTMG) longwave and shortwave radiation, the revised MM5 Monin Obukhov surface layer scheme, the Noah land surface model and the Yonsei University (YSU) planetary boundary layer scheme; cumulus parameterization is enabled on the 15 km domain and disabled on the 3 km domain.[32-35] Terrain related options, including slope radiation, topographic shading and topographic wind correction, are enabled. Observation nudging is applied to wind, temperature and moisture with internal quality control, so the WRF fields represent a dynamically consistent and observation-constrained regional modelling cycle. The WRF simulations used for prior training are from 2024, whereas station verification is performed independently using observations from January to March 2025. Thus, no 2025 verification observations are used in prior training, posterior reconstruction or quantile mapping. Although the WRF simulations may reflect information from assimilated observations through the regional modelling cycle, this influence is temporally separated from the independent station verification period.

EC 0.25° forecasts initialized at 00 and 12 UTC from 1 January to 31 March 2025 are used as posterior constraints.[8,36] Forecast lead times from 0 to 120 h are included in the main verification. For posterior sampling, the EC U10 and V10 fields are aligned with the target domain and represented on a 32 by 24 coarse grid. EC 0.1° forecasts over the same period are processed independently and used only as an operational reference.[8,36]

The station archive provides hourly 10 m wind observations and station coordinates. Values with missing data flags greater than or equal to 999000 are excluded, together with stations outside the reconstruction domain. Approximately 1,000 quality controlled stations are available at each valid time.

## Learning the high resolution wind prior

The high resolution wind prior is learned from WRF simulated U10 and V10 fields on the 0.03° grid. Let x0 denote a two channel high resolution vector wind sample. The diffusion model is trained as an unconditional generative prior for terrain organized vector wind fields.[22-24] In the forward process, Gaussian noise is gradually added to x0, and the reverse model learns to remove this noise and recover WRF consistent kilometre scale wind structures. The purpose of this prior is not to predict a future wind field by itself or to learn a deterministic EC to WRF downscaling rule. Instead, it represents the distribution of plausible fine scale U10 and V10 structures that can later be constrained by the operational forecast during posterior reconstruction.

$$q(\boldsymbol{x}_t|\boldsymbol{x}_0) = \mathcal{N}\left(\boldsymbol{x}_t; \sqrt{\bar{\alpha}_t}\boldsymbol{x}_0, (1-\bar{\alpha}_t)I\right), \boldsymbol{x}_t = \sqrt{\bar{\alpha}_t}\boldsymbol{x}_0 + \sqrt{1-\bar{\alpha}_t}\varepsilon, \varepsilon \sim \mathcal{N}(0, I) \quad (2)$$

where t is the diffusion step, beta is the noise variance schedule, alpha is the retained signal fraction, and $\bar{\alpha}_t$ is the cumulative signal fraction:

$$\alpha_t = 1 - \beta_t, \bar{\alpha}_t = \prod_{s=1}^{t} \alpha_s = \prod_{s=1}^{t} (1 - \beta_s) \quad (3)$$

The reverse process is parameterized as

$$p_\theta(\boldsymbol{x}_{t-1}|\boldsymbol{x}_t) = \mathcal{N}\left(\boldsymbol{x}_{t-1}; \mu_\theta(\boldsymbol{x}_t, t), \Sigma_\theta(\boldsymbol{x}_t, t)\right) \quad (4)$$

$$L_{\text{simple}} = \mathbb{E}_{x_0,\varepsilon,t}[\|\varepsilon - \varepsilon_\theta(\boldsymbol{x}_t, t)\|_2^2] \quad (5)$$

where the neural network is trained to predict the added diffusion noise. The implementation uses a learned range variance setting. The model is optimized using an epsilon prediction loss, so that the network learns the distribution of normalized WRF U10 and V10 fields across different noise levels.

Before training and inference, U10 and V10 are normalized using a shared component scale derived from the WRF training set. With the WRF attributes used here, this shared scale is approximately 2.46 m $s^{-1}$. Wind component values are clipped at five times this shared scale and mapped to [-1, 1]:

$$s = \sqrt{\left[\frac{\sigma_U{}^2 + \sigma_V{}^2}{2}\right]}, \tilde{x} = \frac{\text{clip}\left(\frac{x}{s}, -k, k\right)}{k}, k = 5 \quad (6)$$

The implementation uses a two channel U10 and V10 diffusion model with 1000 diffusion steps and an epsilon prediction objective. The network has 128 base channels, two residual blocks and four attention heads, with learned range variance and attention at resolutions 32 and 16. Training uses a learning rate of $1 \times 10^{-4}$, weight decay of $1 \times 10^{-4}$, loss second moment schedule sampling and an exponential moving average rate of 0.9999. The exponential moving average weights at 360,000 training steps are used as the high resolution prior during posterior reconstruction.

### ECMWF constrained posterior reconstruction

During inference, an EC 0.25° forecast is used as a coarse constraint on the unknown high resolution wind field.[25-31] The constraint is imposed only after the generated field is mapped to the EC scale, so it preserves the large scale circulation and temporal evolution resolved by the operational forecast.[25,26] The WRF trained prior supplies the unresolved kilometre scale wind variability needed to produce terrain aware local structures.

A coarse grid observation operator A maps a 256 × 192 high resolution U10 and V10 field to the 32 × 24 EC scale grid using 8× area averaging:

$$\mathcal{A}(\boldsymbol{x}) = \mathrm{AvgPool}_8(\boldsymbol{x}) \tag{7}$$

At each reverse diffusion step, the model predicts the noise component in the current sample and converts it into a clean high resolution wind estimate. The coarse forecast constraint is then evaluated against this clean estimate rather than against the noisy sample:

$$\widehat{\boldsymbol{x}}_0 = \widehat{\boldsymbol{x}}_0(\boldsymbol{x}_t, t) = \frac{1}{\sqrt{\bar{\alpha}_t}}\left(\boldsymbol{x}_t - \sqrt{1-\bar{\alpha}_t}\boldsymbol{\varepsilon}_\theta(\boldsymbol{x}_t, t)\right) \tag{8}$$

The residual between the EC forecast and the coarse grid representation of the predicted clean field is:

$$\boldsymbol{r}_t = \boldsymbol{y} - \mathcal{A}(\widehat{\boldsymbol{x}}_0) \tag{9}$$

The reverse update is then guided by the gradient of the squared residual:

$$\boldsymbol{x}_{t-1} \leftarrow \boldsymbol{x}'_{t-1} - \lambda \nabla_{\boldsymbol{x}_t} \|\boldsymbol{y} - \mathcal{A}(\widehat{\boldsymbol{x}}_0)\|_2^2 \tag{10}$$

where the $\boldsymbol{x}'_{t-1}$ denotes the unconditional reverse diffusion update, and the conditioning scale $\lambda$ is set to 0.7. This parameter controls the balance between consistency with the EC 0.25° forecast at resolved scales and freedom for the WRF trained prior to restore terrain scale variability. A stronger conditioning term would more tightly enforce the coarse forecast but suppress fine scale structures, whereas a weaker term would allow more prior driven variability but could reduce consistency with the operational forecast. The value used here was chosen as a stable compromise for all experiments and was kept fixed across forecast times and events. Systematic sensitivity to this parameter is left for future applications.

For each EC 0.25° forecast file and valid time, ten independent posterior members are generated; each member initializes a high resolution two channel wind field from Gaussian noise with shape $2 \times 256 \times 192$. The diffusion sampler uses 1000 reverse steps, a cosine noise schedule, epsilon prediction, learned range variance, clipping of denoised values, and no class conditioning. Each posterior member is generated independently under the same EC 0.25° constraint.

## Baselines and verification

Four reference products are used. EC 0.25° is the coarse forecast used during posterior sampling. EC 0.1° is included as a higher resolution operational reference. EC 0.25°-Interp is constructed by bilinearly interpolating EC 0.25° to the 0.03° target grid. EC 0.25°-QM is included as a retrospective distribution adjustment benchmark. It applies empirical quantile alignment to wind speed using a single global space-time mapping from January to March (JFM) 2025 EC 0.25°-Interp to the JFM 2024 WRF training distribution, while preserving the EC wind direction. Because this mapping uses the full evaluation period EC distribution, EC 0.25°-QM is not intended as a real-time operational baseline.

### Station verification

All gridded products are matched with station observations at the corresponding valid times using nearest neighbour grid cell sampling at station coordinates; bilinear interpolation is used only to construct the EC 0.25°-Interp grid before station sampling. Only samples that are simultaneously available for all products are retained. Forecast accuracy is measured using RMSE and the Pearson correlation coefficient over forecast lead times from 0 to 120 h. Because station-hour samples are not independent, skill differences are interpreted as descriptive statistics over matched station and forecast samples rather than as independent sample significance tests. Robustness is further assessed using event wise verification and additional MAE diagnostics.

### Terrain analysis

Terrain complexity is quantified using local relief within a 5 km radius of each station. Local relief is defined as the difference between the 90th and 10th percentiles of terrain elevation within the neighbourhood. Stations are divided into four classes according to the quartiles of this metric.

### Strong wind event selection

Strong wind events are identified from hourly station observations between January and March 2025. Consecutive periods are treated as separate events when their peak times are separated by at least 24 h. An event is retained when at least three stations record wind speeds of 12 m $s^{-1}$ or greater. The peak hour of each eligible event is used for event based verification, resulting in 13 events.

## Data availability

The ECMWF forecast products used as EC 0.25° and EC 0.1° are available through ECMWF data services (https://www.ecmwf.int/en/forecasts/datasets; https://data.ecmwf.int/forecasts) subject to the applicable ECMWF terms of use. ERA5 and ERA5-Land data are available from the Copernicus Climate Data Store (https://cds.climate.copernicus.eu). The WRF model simulations generated and analysed during this study and the processed KiloGen training and testing datasets are not publicly available because of data storage limitations, but are available from the corresponding author on reasonable request. The station observations used for evaluation are subject to third party data restrictions and are available from the corresponding data providers; processed station based evaluation data may be available from the corresponding author on reasonable request and with permission of the data providers.

## Code availability

The KiloGen code is implemented in PyTorch, a Python-based deep learning framework available at https://github.com/pytorch/pytorch. The source code required to train the WRF based wind prior, perform ECMWF-constrained posterior sampling and conduct station verification is available for peer review at https://github.com/cyjsss999/KiloGen. The repository will be made publicly available upon acceptance. The original WRF simulations, ECMWF forecasts and station observations are not redistributed because of data volume and data use restrictions. The repository provides data schema documentation and synthetic input generators to support workflow reproducibility.

## Acknowledgements

This research is supported by Smart-Grid National Science and Technology Major Project (Grant No. 2025ZD0805500). The authors thank ECMWF for providing the ERA5 reanalysis data and ECMWF forecast products, and the China Meteorological Administration for providing the surface meteorological observation data.

During manuscript preparation, ChatGPT was used to assist with language editing and formatting checks. All scientific content, analyses, interpretations and conclusions were developed, verified and approved by the authors.

## Author contributions

Y.J.C. and Y.W. conceived the study. Y.J.C. developed KiloGen, processed the data, performed the experiments, carried out the verification and visualization, and drafted the manuscript. Y.W. provided scientific guidance throughout the study, contributed to the methodological design and experimental strategy, helped refine the research ideas, and revised the manuscript. S.M.F. supervised the research, guided the overall research direction, supported the project framework and funding acquisition. S.L.F. and B.W. coordinated the renewable energy project collaboration, provided the WRF simulation data. H.Y. contributed to the electric power application context and provided station observation data. H.J.L. contributed to methodological discussions and manuscript review. Y.M.L. and X.Y.S. contributed to data processing, analysis and manuscript review. All authors reviewed and approved the final manuscript.

## Competing interests

All authors declare no financial or non-financial competing interests.

Supplementary Information for

# Zero Shot Kilometre Scale Wind Forecasting over Wind Rich Complex Terrain using Diffusion Posterior Sampling

Yujiang Cai[1,2], Ya Wang[1,2], Shuanglei Feng[3], Bo Wang[3], Yiming Liu[1,2], Hui Yuan[4], Xinyi Sun[1,2], Haijie Li[1,2] and Shenming Fu[1,2]

*[1] China Key Laboratory of Earth System Numerical Modeling and Application, Institute of Atmospheric Physics, Chinese Academy of Sciences, Beijing, 100029, China.*

*[2] University of Chinese Academy of Sciences, Beijing, 100049, China.*

*[3] National Key Laboratory of Renewable Energy Grid-Integration, China Electric Power Research Institute, Beijing 100192, China.*

*[4] State Grid Shanxi Electric Power Research Institute, Taiyuan 030001, China.*

Corresponding author: Dr. Ya Wang (wangya@mail.iap.ac.cn)

Dr. Shenming Fu (fusm@mail.iap.ac.cn)

**Contents of this file**

**Supplementary Fig. 1 | Wind-direction verification and station-scale direction errors.** a, Lead-time RMSE of 10-m wind direction for EC 0.25°, EC 0.1° and KiloGen 0.03°. b, Lead-time direction consistency, measured as the mean cosine of the angular difference between modelled and observed wind direction. c-e, Station wind-direction RMSE maps for EC 0.25°, EC 0.1° and KiloGen 0.03°, respectively. Colours denote station RMSE in degrees. Percentages in e indicate the domain-mean RMSE reduction of KiloGen relative to the EC baselines.

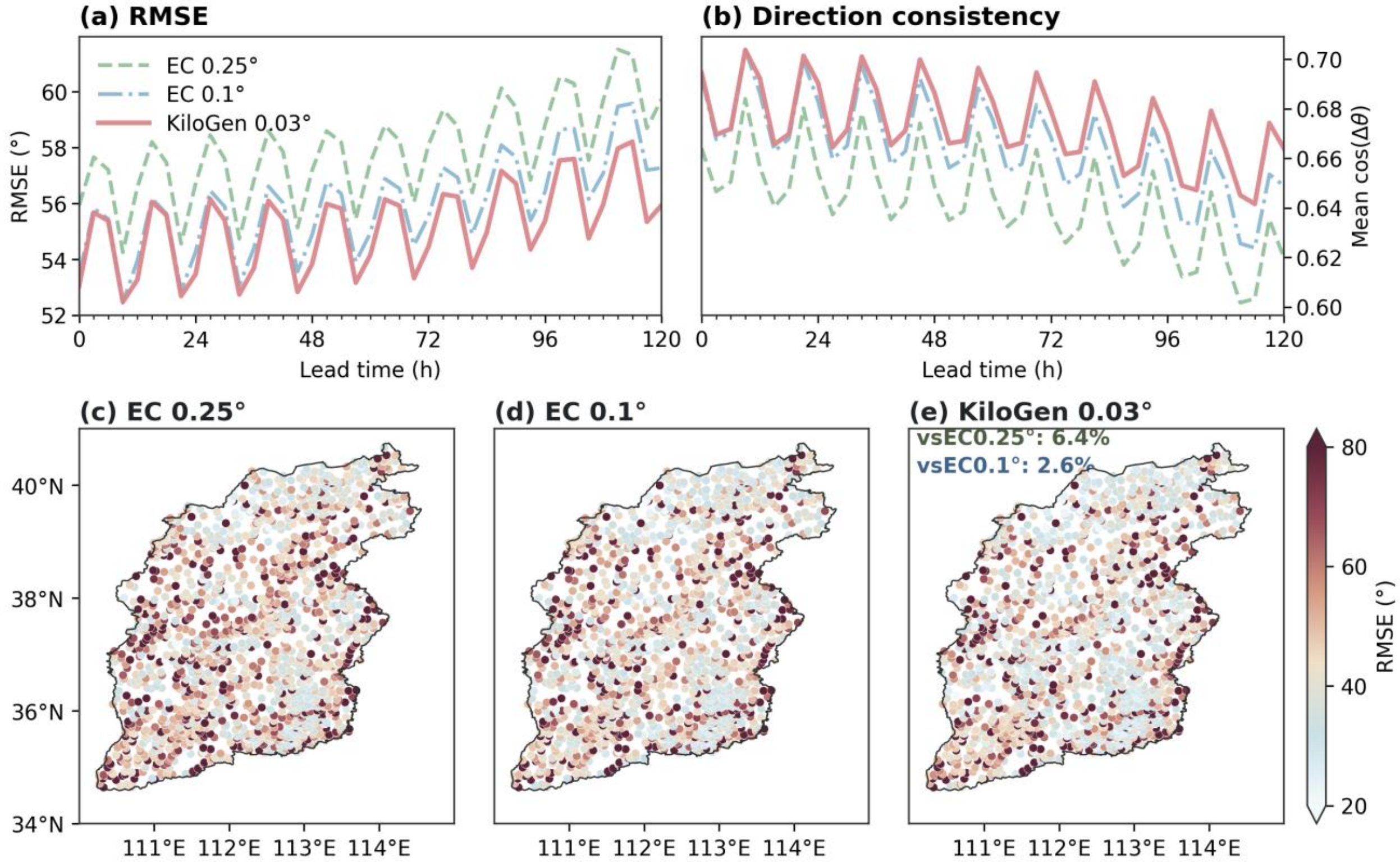

**Supplementary Fig. 2 | Additional deterministic verification metrics.** a, Lead-time MAE of 10-m wind speed from 0 to 120 h for EC 0.25°, EC 0.25°-Interp, EC 0.25°-QM, EC 0.1° and KiloGen 0.03°. b, Event-wise strong-wind MAE for the independent strong-wind events. The metrics provide an additional deterministic check that complements the RMSE and correlation diagnostics in the main text.

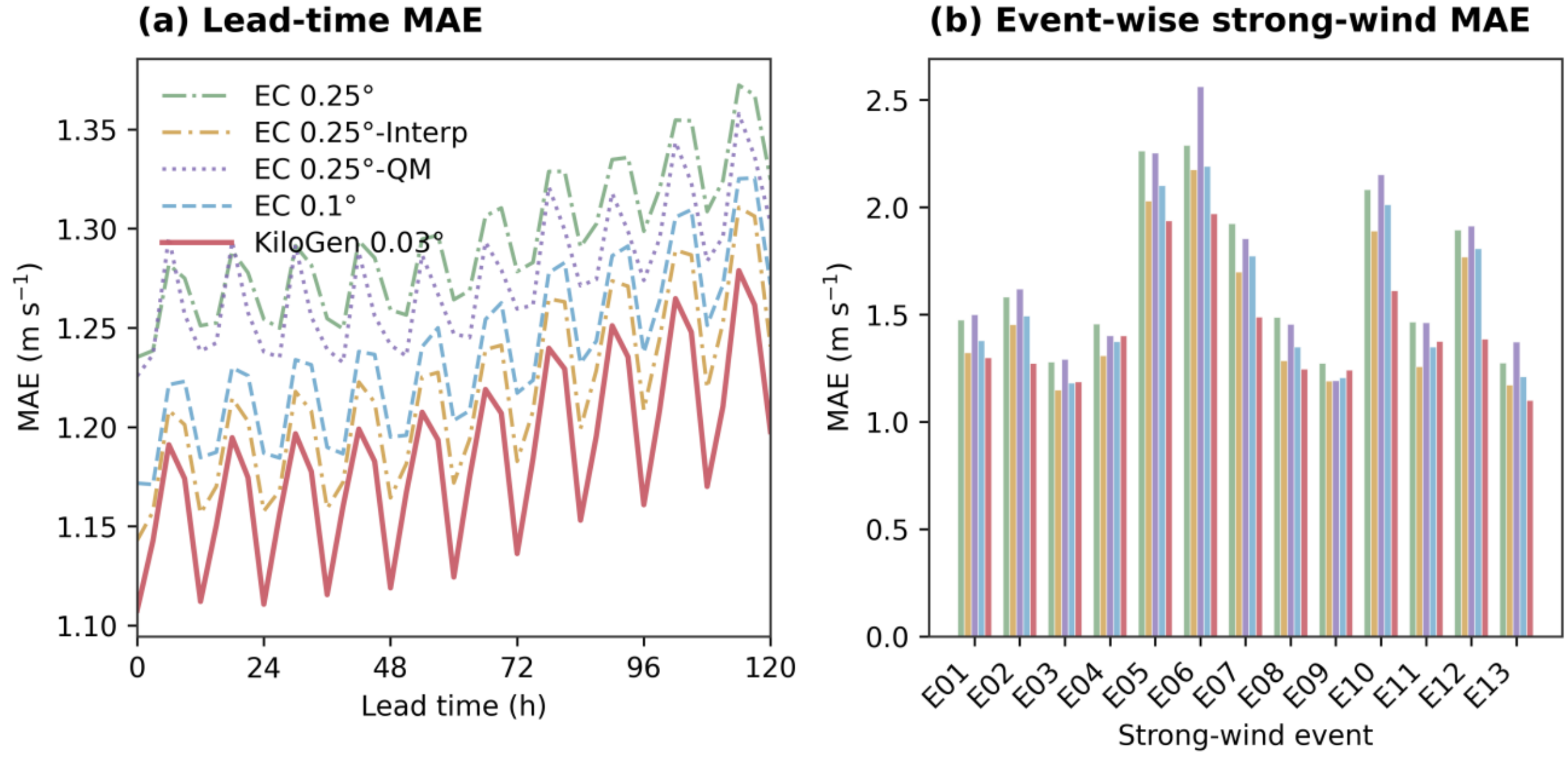

**Supplementary Fig. 3 | Catalogue of independent strong-wind events.** The catalogue shows the independent strong-wind events identified from hourly station observations during January-March 2025. Events are separated by at least 24 h, and the peak hour of each eligible event is retained when at least three stations record wind speed ≥12 m $s^{-1}$. Marker attributes follow the same event definition used in the main-text multi-event verification.

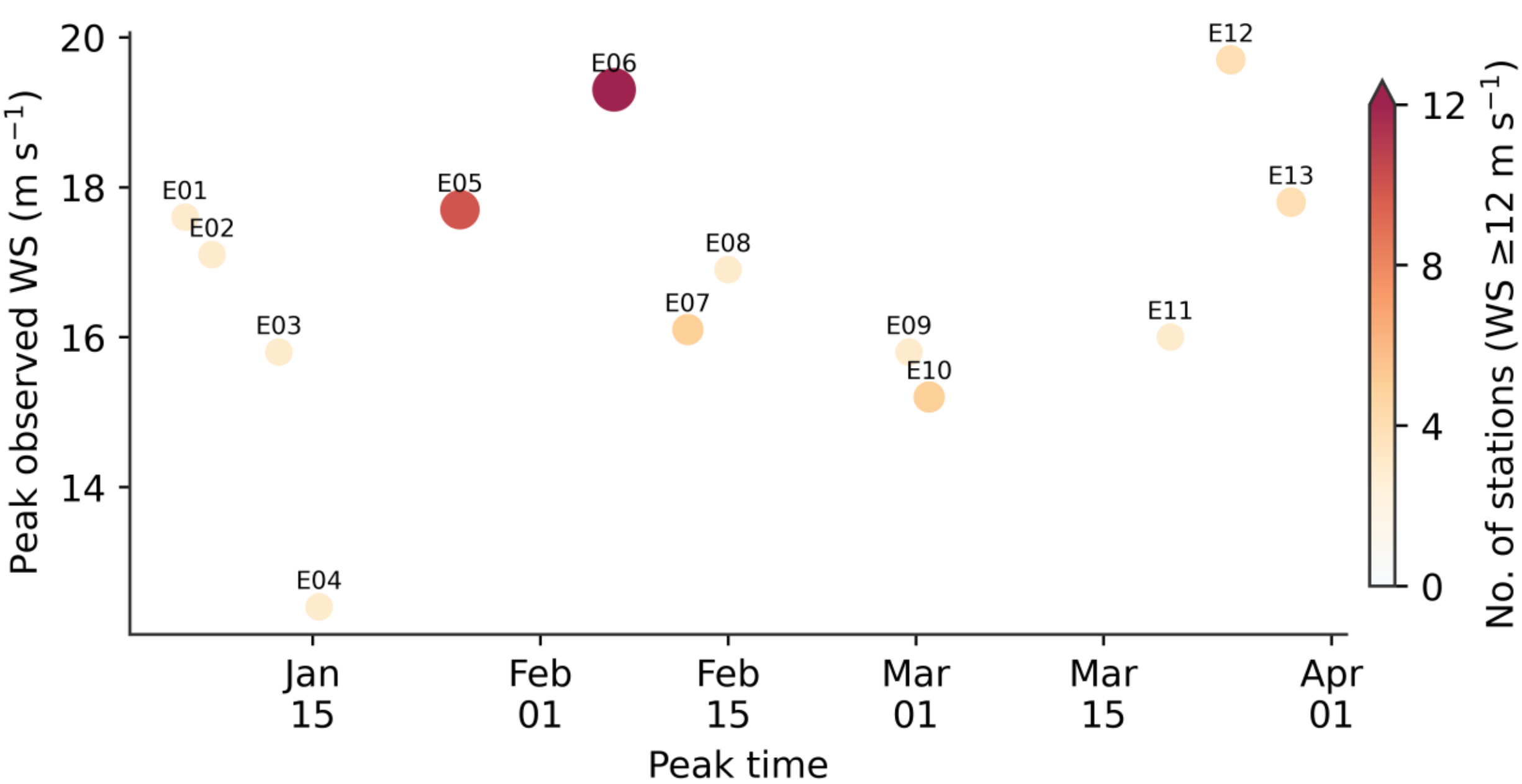

**Supplementary Fig. 4 | Extended examples of strong-wind case reconstructions.** Rows show three additional strong-wind events: E05, E10 and E12. The left column shows EC 0.25°, and the right column shows KiloGen 0.03° for the same valid times. Shading denotes 10-m wind speed, arrows denote 10-m horizontal wind vectors, and station markers indicate observed strong-wind sites. These examples extend the representative E06 case in the main text and show the event-to-event consistency of the terrain-aware reconstruction.

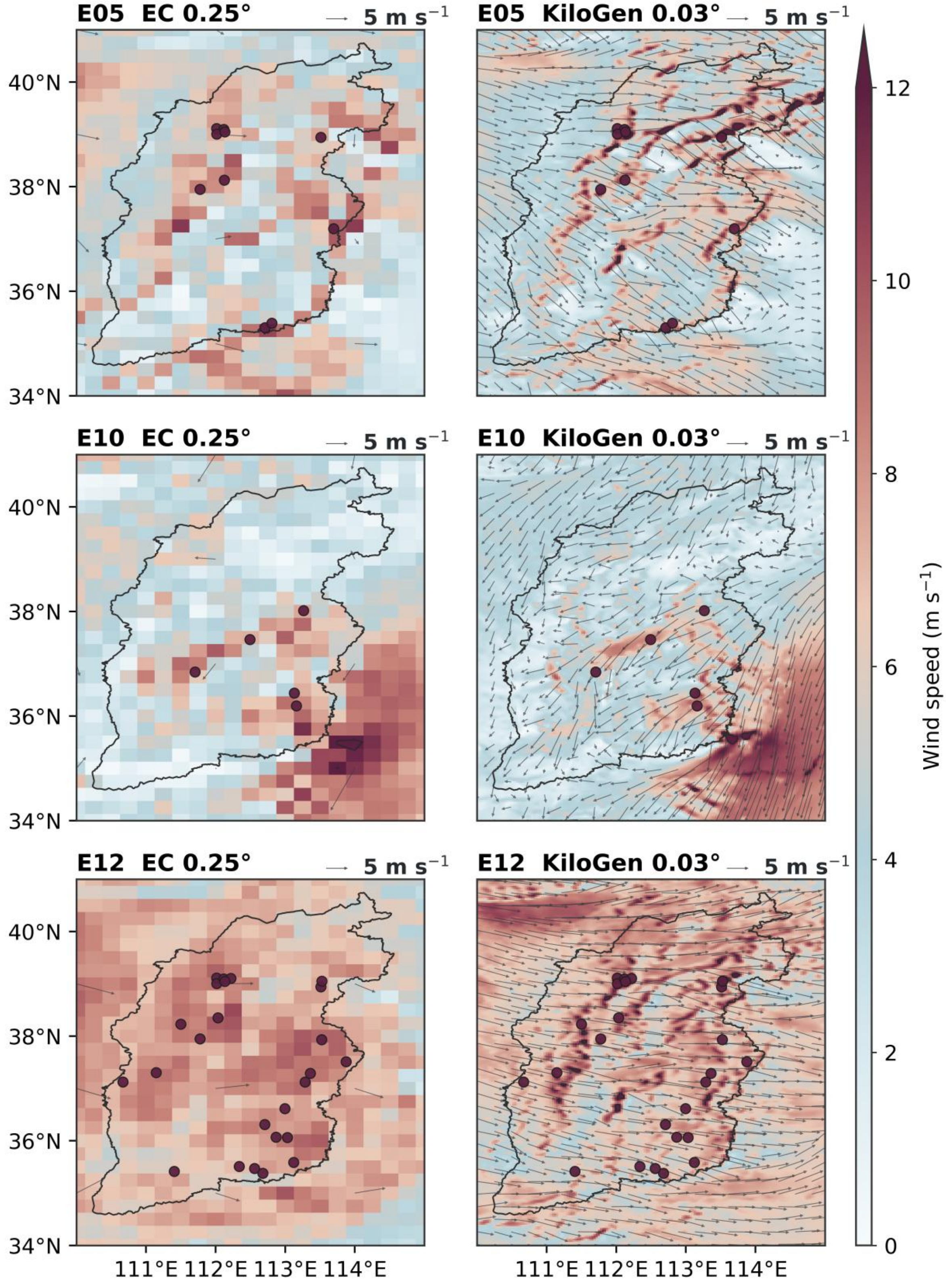

**Supplementary Fig. 5 | Posterior members and ensemble spread for the E06 strong-wind case.** a, EC 0.25° forecast. b, KiloGen ensemble mean. c-e, Three individual KiloGen posterior members. f, Posterior ensemble spread of 10-m wind speed across the 10 KiloGen members. Shading in a-e denotes wind speed, arrows denote horizontal wind vectors, and shading in f denotes spread. Station markers indicate observed strong-wind sites at 03 UTC on 6 February 2025. The member fields illustrate the range of plausible kilometre-scale reconstructions allowed by the WRF-trained prior under the same EC 0.25° constraint.

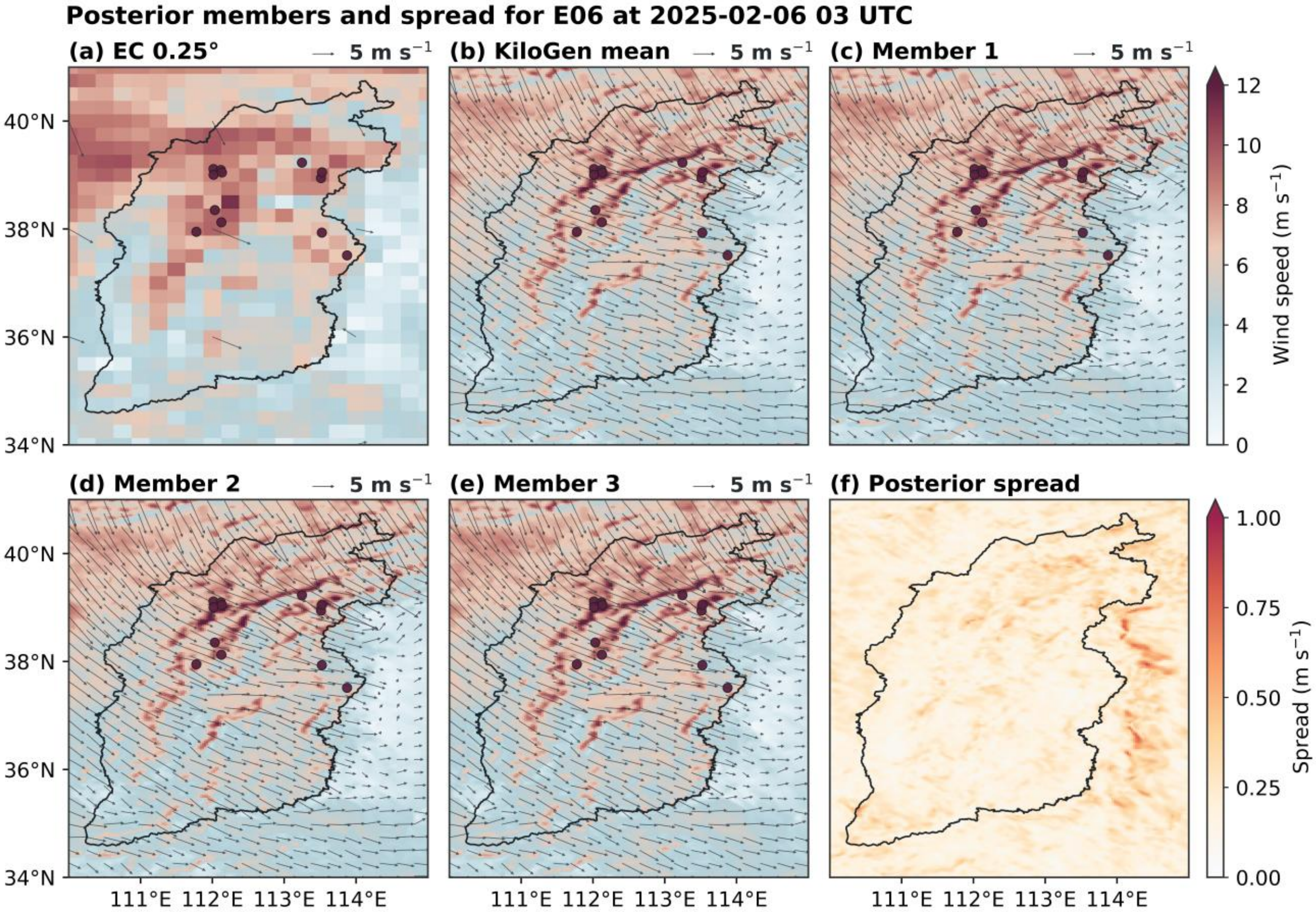